\pdfoutput=1

\documentclass{appolb}
\usepackage{graphicx}
\usepackage[comma,square,numbers,sort&compress]{natbib}
\usepackage{lineno}

\def\pt{\mbox{$p_{\rm T} $}} 
   
\def\snn{\mbox{$\sqrt{s_{\rm NN}}$}}   
\def\pb {Pb\mbox{--}Pb }

\def\j {\ensuremath{\mathrm{J}\kern-0.02em/\kern-0.05em\psi} }
\def\jwospace {\ensuremath{\mathrm{J}\kern-0.02em/\kern-0.05em\psi}}

\def\raa {\mbox{$R_{\rm{AA}}$}}

\newcommand{ \be }{\begin{eqnarray}}
\newcommand{ \ee }{\end{eqnarray}}

\newcommand {\cent}[2]      {\ensuremath{\mathrm{#1\mbox{--}#2\%}}}
\newcommand {\Npart}      {\ensuremath{\langle N_{\mathrm{part}} \rangle}}
\newcommand {\mevc}      {\ensuremath{\mathrm{MeV}\kern-0.05em/\kern-0.02em c}}
\newcommand {\mevcc}      {\ensuremath{\mathrm{MeV}\kern-0.05em/\kern-0.02em c^2}}
\newcommand {\gevc}      {\ensuremath{\mathrm{GeV}\kern-0.05em/\kern-0.02em c}}
\newcommand {\gevcc}     {\ensuremath{\mathrm{GeV}\kern-0.05em/\kern-0.02em c^2}}

\begin{document}
\title{Measurement of an excess in the yield of J/$\psi$ at very low-$p_{\rm T}$ in Pb--Pb collisions with the ALICE detector%
\thanks{Presented at EDS Blois 2015 Conference, Borgo, Corsica}%
}
\author{L. Massacrier\footnote{Now at LAL, Universit\'e Paris-Sud, CNRS/IN2P3, Orsay, France and Institut de Physique Nucl\'eaire d'Orsay (IPNO), Universit\'e Paris-Sud, CNRS/IN2P3, Orsay, France.} for the ALICE Collaboration   \address{Universit\'e de Nantes, Ecole des Mines and CNRS/IN2P3, Nantes, France 
}
\\
}
\maketitle


\begin{abstract}
We report on the measurement of J/$\psi$ production at very low transverse momentum ($p_{\rm T} < $ 300 MeV/$c$) in Pb--Pb collisions performed with the ALICE detector at the LHC. We find an excess in the yield of J/$\psi$ with respect to expectations from hadronic production. Coherent photo-production of J/$\psi$ is proposed as a plausible origin of this excess. We show the nuclear modification factor of very low-$p_{\rm T}$ J/$\psi$ as a function of centrality. Then we measure the J/$\psi$ coherent photoproduction cross section in peripheral events assuming that it is the mechanism at the origin of the measured excess. It's worth noting that the observation of J/$\psi$ coherent photoproduction in Pb--Pb collisions at impact parameters smaller than twice the nuclear radius has never been observed so far and would open new theoretical challenges. 
\end{abstract}
\PACS{25.75.-q, 25.75.Cj 13.25.Gv, 13.40.-f} 
  
\section{Introduction}

Ultra-relativistic heavy-ion collisions are tools to study the nuclear matter at high temperature and pressure where the formation of the Quark Gluon Plasma (QGP) occurs. Open heavy flavours and quarkonia are interesting probes of the QGP \cite{Andronic:2015wma} since they are produced at the early stages of the collision and might interact with the deconfined medium. In the following, we report on the measurement of J/$\psi$ production at very low-$p_{\rm T}$ ($p_{T} < $ 300 MeV/c) in hadronic Pb--Pb collisions at $\sqrt{s_{NN}}$ = 2.76 TeV. We found an excess in the yield of J/$\psi$ mesons whose magnitude cannot be explained by any of the predicted hot medium effects affecting quarkonium production (such as recombination of charm quarks during \cite{Zhao:2011cv,Liu:2009nb,Thews:2000rj} or at the end of the deconfined phase \cite{BraunMunzinger:2000px,Andronic:2011yq}). Coherent photoproduction of J/$\psi$ in Pb--Pb collisions with impact parameters smaller than two times the nuclear radius (b$<$2R) is proposed as the physics mechanism at the origin of the observed excess. Coherent photoproduction of J/$\psi$ is well known in Ultra-Peripheral Collisions (UPC) at b$>$2R and has been measured by ALICE at the LHC \cite{Abelev:2012ba,Abbas:2013oua}. In UPC, one of the Pb nuclei is a photon-emitter whose interaction with the gluon field of the other nuclei permits to produce a J/$\psi$ at very low-$p_{T}$. Indeed, the coherence condition imposes a maximum transverse momentum for the produced J/$\psi$ of the order of one over the nuclear radius. Coherent photoproduction of J/$\psi$ has never been observed for b$<$2R and the survival of the coherence requirement during an hadronic interaction would open theoretical challenges.

\section{Experimental setup}

The ALICE muon spectrometer permits to detect quarkonium at forward rapidity via their $\mu^{+}\mu^{-}$ decay channel and down to~\pt=0. It consists of a 10 interaction length thick absorber, a 3 Tm dipole magnet, five tracking stations and two triggering stations. Other relevant detectors for this analysis are the Silicon Pixel Detector (SPD), the scintillator arrays (V0) and the Zero Degree Calorimeters (ZDC). The SPD reconstructs the primary vertex. The background induced by the beam and electromagnetic processes is reduced by the V0 and ZDC timing information and by requiring a minimum energy deposited in the neutron ZDC. A dimuon opposite-sign trigger ($\mu\mu$MB) was used in this analysis. It requires at least one pair of opposite-sign track segments in the muon trigger with a \pt~above the 1 GeV/c threshold of the online trigger algorithm, in coincidence with a minimum bias (MB) trigger. The MB trigger is fired by a signal in the V0 detectors at forward and backward rapidity. The data sample analyzed amounts to $L_{int} \sim$ 70 $\mu b^{-1}$. Only the 90$\%$ most central Pb--Pb collisions, for which the MB trigger is fully efficient, are kept. Centrality classes, average number of participant nucleons $\ensuremath{\langle N_{\mathrm{part}} \rangle}$ and average nuclear overlap function \ensuremath{\langle T_{\mathrm{AA}} \rangle} are obtained from a Glauber calculation \cite{Miller:2007ri}.    
More details about the ALICE detector can be found in \cite{Aamodt:2008zz}. 

\section{Analysis and results}
\begin{figure}[t]
{\centering 
\resizebox*{0.7\columnwidth}{!}{\includegraphics{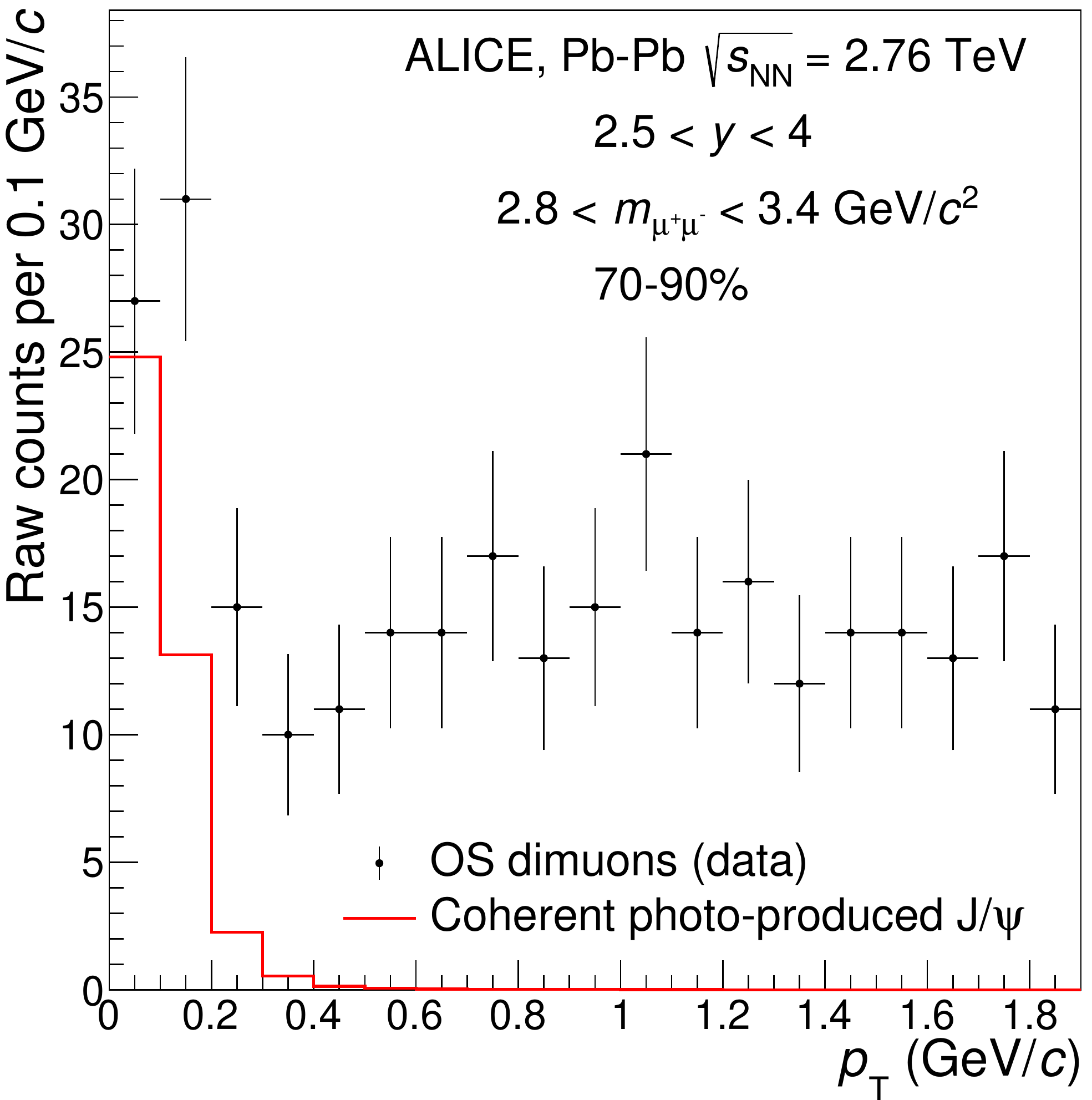}}
\par}
\caption{(Color online) Raw OS dimuon \pt~distribution for the invariant mass range 2.8~$<$~$m_{\mu^+\mu^-}$~$<$~3.4~\gevcc~and centrality class \cent{70}{90}. The red line represents the \pt~distribution of coherently photo-produced J/$\psi$ as predicted by the STARLIGHT MC generator \cite{starlightMC} in \pb ultra-peripheral collisions and convoluted with the response function of the muon spectrometer.}
\label{Fig:ptdistrib}
\end{figure}

Muons reconstructed in the geometrical acceptance of the muon spectrometer and matching a trigger track segment above the 1 \gevc~\hspace{0.2 true cm}\pt~threshold are combined into opposite sign (OS) pairs to form dimuon candidates. Figure~\ref{Fig:ptdistrib} shows the \pt~distribution of OS dimuons in the invariant mass range 2.8~$<$~$m_{\mu^+\mu^-}$~$<$~3.4~\gevcc~and centrality class 70-90$\%$, where the J/$\psi$ contribute to a large fraction of the dimuon yield. A large excess is clearly seen at very low-$\pt$, before any background subtraction. Its \pt~shape follows the trend of the coherently photoproduced J/$\psi$ one in Pb--Pb UPC, obtained by the STARLIGHT MC generator~\cite{starlightMC}.

A way to quantify this excess is to measure the J/$\psi$ nuclear modification factor \raa~defined as:

\begin{equation}  
\raa = \frac{Y^{Pb-Pb}_{J/\psi}}{\langle T_{\rm AA} \rangle \times \sigma^{pp}_{J/\psi}} 
\hspace{-0.5 true cm}
\qquad
\mbox{with \hspace{0.2 true cm}}
Y^{Pb-Pb}_{J/\psi} = \frac{N_{J/\psi}}{BR_{J/\psi \rightarrow \mu^{+}\mu^{-}} \times N_{MB} \times A\epsilon},
\end{equation} 

\begin{figure}[t]
{\centering 
\resizebox*{0.8\columnwidth}{!}{\includegraphics{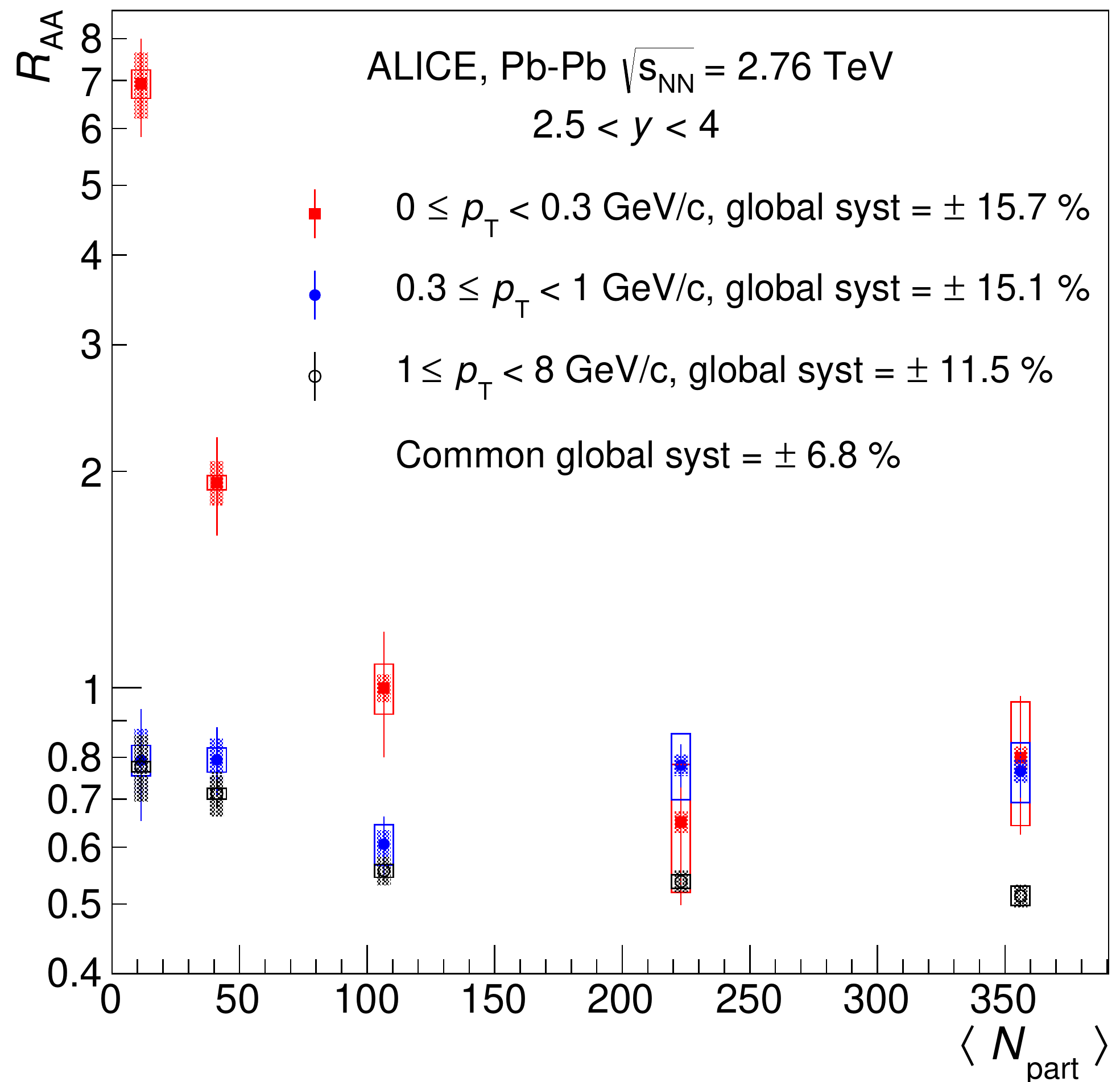}}
\par}
\caption{(Color online) \j \raa~as a function of \Npart~for 3 \pt~ranges in \pb collisions at \snn~= 2.76 TeV. Vertical bars represent the statistical uncertainties. Open boxes are uncertainties uncorrelated in \pt~and centrality. Shaded areas are uncertainties correlated in \pt~but not in centrality. Global systematics quoted in the legend for each \pt~range are the uncertainties correlated as a function of centrality but not in \pt. Common global systematics in the legend are correlated uncertainties. More details about the sources of systematic uncertainties can be found in \cite{Adam:2015gba}.}
\label{Fig:RAA}
\end{figure}

\noindent{
where $Y^{Pb-Pb}_{J/\psi}$ is the J/$\psi$ corrected yield in Pb--Pb in a given range in \pt~and centrality,  $\langle T_{\rm AA} \rangle$ is the mean nuclear overlap function in a given centrality class, and $\sigma^{pp}_{J/\psi}$ is the J/$\psi$ pp reference cross section in a given \pt~range. $N_{J/\psi}$ is the extracted J/$\psi$ raw yield , $BR_{J/\psi \rightarrow \mu^{+}\mu^{-}}$ the branching ratio in the dimuon decay channel, $N_{MB}$ the number of minimum bias events and $A\epsilon$ the J/$\psi$ acceptance times efficiency assuming hadroproduction. To evaluate the J/$\psi$ \raa, a similar procedure as in \cite{Abelev:2013ila} was followed except for the determination of the J/$\psi$ pp reference cross section. In the \pt~range 1--8~\gevc, the J/$\psi$ cross section in pp collisions at $\sqrt{s}$ = 2.76 TeV is taken from the ALICE measurement \cite{Aamodt:2011gj} while in the low-\pt~ranges, due to limited statistics, it is obtained (as well as the corresponding systematic uncertainty) by fitting the \pt-differential cross section distribution with the parametrizations in \cite{Bossu:2011qe},\cite{Tsallis:1987eu} and in \cite{Albajar:1989an}. In addition, the validity of the procedure was confirmed using the \j data sample in pp collisions at 7 TeV \cite{Abelev:2014qha}.
Figure \ref{Fig:RAA} shows the J/$\psi$ \raa~in three \pt~ranges (0--0.3, 0.3--1, 1--8 \gevc) and five centrality classes (0--10\%, 10--30\%, 30-50\%, 50-70\%,70-90\%). A strong increase of the \raa~is observed in the very low-\pt~range for the most peripheral Pb--Pb collisions, reaching a value up to about 7 in the centrality range 70--90\%. Such a behaviour is not predicted by models which include a regeneration component\cite{Zhao:2011cv,Liu:2009nb,Thews:2000rj}.}

The excess of J/$\psi$ at very low-\pt~can also be quantified by subtracting the number of J/$\psi$ expected from hadroproduction in Pb--Pb to the measured number of J/$\psi$. Details about the evalutation of the number of expected J/$\psi$ from hadroproduction in a data driven way can be found in \cite{Adam:2015gba}. The significance of the excess is 5.4$\sigma$, 3.4$\sigma$ and 1.4$\sigma$ in the 70--90\%; 50--70\%, 30-50\% centrality ranges, respectively. Assuming that coherent photoproduction of J/$\psi$ for b $<$ 2R is at the origin of the excess, the corresponding cross section can be obtained, following a similar procedure as in \cite{Abelev:2012ba}. Results are reported in table \ref{tab:jpsiexcess}. The cross section per unit of rapidity in the centrality class 70--90\% amounts to 59 $\pm$ 11 $^{+7}_{-10}$ $\pm$ 8 ($\mu$b). Coherent photoproduction of J/$\psi$ in peripheral hadronic Pb--Pb collisions was not predicted, therefore no theoretical calculations existed at the time of the measurement. To have a rough estimate of the cross section, we considered the extreme asumption that all the charges in the source ion and all nucleons in the target ion contribute to the photonuclear cross section as in coherent UPC. We obtained a cross section of about 40 $\mu$b which is of the same order of magnitude as our measurement. Recent calculations triggered by this presentation \cite{Klusek-Gawenda:2015hja}, with a different way to treat the overlap region, tends to favour a scenario in which only the spectator region contributes to the coherent photoproduction.

\begin{table*}[t]
\begin{center}
\begin{tabular}{|c|c|c|c|c|c|} \hline
\scriptsize{Centrality (\%)} & \scriptsize{0--10} & \scriptsize{10--30} & \scriptsize{30--50} & \scriptsize{50--70} & \scriptsize{70--90} \\ \hline
\scriptsize{d$\sigma_{\j}^{\rm coh}/$d$y$  ($\mu$b)} & \scriptsize{$<$ 318} & \scriptsize{$<$ 290} & \scriptsize{73 $\pm$ 44 $^{+26}_{-27}$ $\pm$ 10} & \scriptsize{58 $\pm$ 16 $^{+8}_{-10}$ $\pm$ 8} & \scriptsize{59 $\pm$ 11 $^{+7}_{-10}$ $\pm$ 8} \\ 
\hline
\end{tabular} 
\end{center}

\caption{\j coherent photoproduction cross section per unit of rapidity in five centrality classes, in \pb collisons at \snn~= 2.76 TeV quoted with its statistical, uncorrelated and correlated systematic uncertainties respectively. More details about the sources of systematic uncertainties can be found in \cite{Adam:2015gba}. In the most central classes, an upper limit (95$\%$ CL) on the cross section is given.}
\label{tab:jpsiexcess}
\end{table*}

\section{Conclusion and outlook}

In summary, the ALICE Collaboration has observed a J/$\psi$ yield enhancement at  very low-\pt~(0--0.3~\gevc) in peripheral Pb--Pb collisions at $\sqrt{s_{NN}}$ = 2.76~TeV. Remarkably, the J/$\psi$ nuclear modification factor reaches about 7 in the centrality class 70--90\%. Coherent photoproduction of J/$\psi$ occuring at impact parameters smaller than twice the nuclear radius is proposed as the underlying physics mechanism at the origin of the excess. A good qualitative agreement is obtained between the measured photoproduction cross section and UPC calculations extrapolated to peripheral Pb--Pb collisions. However, 
this measurement remains a challenge for theoretical studies since the implementation of coherent photoproduction in nuclear collisions is complex and all the aspects should be accounted for. If confirmed coherent photoproduction of J/$\psi$ at b $<$ 2R could potentially become a powerful new QGP probe since those J/$\psi$ would be formed at the early stages of the collision, with well defined kinematics, and could be dissociated by color screening in the QGP. A theoretical model is therefore essential to provide a reference for this probe in the absence of QGP formation. During Run 2 (2015-2019) at the LHC, the measurement of the J/$\psi$ very low-\pt~excess will be extended to semi-central and central collisions.

\begin{scriptsize}
\bibliographystyle{science} 
\bibliography{template}
\end{scriptsize}

\end{document}